\documentclass[12pt,a4paper]{article}
\usepackage[]{epsfig}	      
\begin{document}
 
 \thispagestyle{empty}
 
 \title{Surface Embedding, Topology and Dualization for Spin Networks.}

 \author{Peter Kramer$^a$ and Miguel Lorente$^b$,\\
 \small $^a$ Institut f\"ur Theoretische Physik der 
 Universit\"at D 72076 T\"ubingen, Germany,\\
 \small $^b$ Departamento de Fisica, Universidad de Oviedo, E 33007 Oviedo, Spain.}
 \date{May 15, 2002}
 \maketitle

\section*{Abstract.}
Spin networks are graphs derived from $3nj$ symbols of
angular momentum. The surface embedding, the topology and
dualization of these networks are considered. Embeddings
into compact surfaces include the orientable sphere $S^2$ 
and the torus $T$, and the not orientable projective space $P^2$ and
Klein's bottle $K$. Two families of $3nj$ graphs admit
embeddings of minimal genus into $S^2$ and $P^2$. Their
dual 2-skeletons are shown to be triangulations of these surfaces.

\section{Spin networks.}

Spin networks were introduced by Penrose \cite{PE}
as a purely combinatorial description of the geometry of 
spacetime.
A spin network has a graph $\Gamma$ of $3n$ edges
which represents an invariant formed from 
$3n$  angular momenta $j$. $\Gamma$  is connected and 
at any vertex has degree $3$.
When $j$-values are attached to $\Gamma=\Gamma(3nj)$ which 
fulfill the intertwining rules at each vertex, the value
of the corresponding $3nj$ invariant can be computed and used in 
the theory of spin networks. A systematic description and 
evaluation of the $3nj$ invariants is given by Yutsis et al.
in \cite{YU}. 
Motivated by the discretization of a Riemannian surface to 
avoid infinities in Einstein action, Ponzano and Regge \cite{PON}
extensively used $3nj$ symbols and their geometrical interpretation.
Hasslacher and Perry \cite{HAS} proved that Ponzano Regge discretization
is a particular case of spin networks. The development of
this subject is reviewed in \cite{REG}. The spin network approach uses 
combinatorial properties of a set of relations to give an interpretation
of the structure of space. In the present paper we explore a similar
correspondence between combinatorial properties of graphs and their
embedding in compact surfaces. Our method is in line with algebraic
topology and provides a more rigorous setting for the
intuitive embedding of $3nj$ symbols given by Ponzano and Regge.
The philosophical background to a spin network interpretation goes back to
the relational theory of space and time proposed by Leibniz.
One of the authors (M.L.) has developed in  \cite{LO1},\cite{LO2} 
a fundamental theory based
on Leibniz' ideas to give an interpretation of discrete 
models of space and time.

\section{Topology of spin networks.}

We turn to the topological aspects of spin networks.
Underlying a spin network
is a graph $\Gamma$. If its vertices are taken as points and its edges as lines,
its geometry provides a 1-skeleton. We shall use this and other  notions of
cell complexes  taken from
algebraic topology \cite{MU}. 
A topological analysis of spin networks in our opinion 
would set the ground for the following explorations:

(1) Ponzano and Regge in their work \cite{PON} of 1968 discuss 
extensively  graphs for $3nj$ symbols and speak about surfaces, 
polytopes and subdivisions associated with them.  
In their approach to discrete quantum gravity they consider the 
$3nj$ symbols, starting from $6j$, as polytopes in $E^3$. Topological 
properties of higher symbols are sketched only for the $9j$ symbol. 
But many of the general spin networks with $3nj$ graphs, beginning with 
$\Gamma(9j)$ as given by Yutsis \cite{YU} p. 60 Fig. 18.1, do not admit an orientable 
embedding into $S^2$
and $E^3$.
It will be necessary to restrict the class of spin networks or else
to find new ways of dealing with their embedding.
Recent mathematical analysis can provide
and classify the topology of $3nj$ graphs and their
variety.
\vspace{0.2cm}

(2) In the discussion of spin networks there appears the notion of
duality, compare for example Baez \cite{BA1}. Since the graphs of 
spin networks have degree $3$
at any vertex, it is conjectured that the dual has
triangular faces. This triangulation is also anticipated as a geometrical 
tool in \cite{PON}. But duality of networks, in the notion of 
algebraic topology \cite{MU},  works only provided 
the graph $\Gamma$ has been embedded into  a surface. So 
embedding must precede dualization and triangulation.
\vspace{0.2cm}

(3) A graph $\Gamma$ seen from its vertex (point)
set $V$ is a combinatorial rule to select pairs of vertices linked by
a set of edges $E$. A natural extension from  the notion  of a graph
to a richer geometry and topology is to consider   
closed sets of edges on $\Gamma$ as candidates for faces.
The analysis of surface embeddings allows to equip 
families of graphs with faces and topological data like Euler characteristics,
orientation and genus obtained in this
process. The universal covering of the embedding surface  
provides a corresponding covering of the embedded graph.
Models for physics with discrete geometry could
take advantage of this access. Regge \cite{REG2} has 
shown how geometric notions such as curvature arise in discrete
geometries. 
\vspace{0.2cm}

In what follows we restrict the analysis of spin networks to
topological questions. We analyze for graphs the embedding, topology
(Euler characteristics $\chi$ and genus $g$),
dualization and possible triangulation  under the following restrictions:\\
(i) The graphs $\Gamma$ will be taken from spin networks.
The corresponding restrictions on the graphs as given in \cite{YU}
are summarized in section 6.\\
(ii) We introduce faces into the graphs of spin networks, but restrict 
their choice such that all of them can be embedded into
a single compact  surface ${\cal S}$. \\
The dimension $d=2$ is sufficient for studying duality and may be taken 
towards $d>2$ in later steps, see section 6.
Compactness excludes in particular embeddings into manifolds
with boundaries. So closure  is in line with the standard  notions of 
space in physics.
The compact surfaces admit a well-known classification of their topology
\cite{SE} which we briefly review in section 3.
They are uniquely characterized by their Euler characteristics,
orientability and genus. Embeddings restricted
to these surfaces become
accessible to mathematical analysis. In section 4 we describe
the theory for the  embedding of  graphs into surfaces. This
theory was given in 1995 by Bonnington and Little \cite{BO},
it was reanalyzed and presented as an algorithm by Schwark \cite{SCH}.  
In section 5 we apply the method to the simplex. We demonstrate
the variety of surfaces including $S^2, T, P^2$ and Klein's bottle $K$, and genus 
ranging as $g=0,1,2,3$. In section 6 we find embeddings of
minimal genus for two families of $3nj$ graphs. 
In section 7 we dualize 
the 2-skeleton for the minimal embeddings 
and find  triangulations of the surfaces $S^2, P^2$.
In the  section 8 conclusion we comment on implications
of topological properties for spin networks.

\section{Compact surfaces and topological polygons.}

A topological surface ${\cal S}$ we take as a compact 2-manifold. This means
that we exclude manifolds with boundaries like the cylinder or the
M\"obius strip. For elementary  concepts and illustrations in the topology
of surfaces we refer to Hilbert and Cohn-Vossen \cite{HI}, sections 44-51,
pp. 290-340.

The topology of compact surfaces can be described by means
of systems of topological polygons, see Seifert and Threlfall \cite{SE}:
Consider in $E^2$  a system ${\cal M}$ of 
$F$ polygons with edges and vertices, each homeomorphic to a unit disc,
with pairwise disjunct point sets. Assume that the number of edges is even and 
that there is a topological map
of each edge and its vertices onto exactly one other edge and its vertices. 
Identify pairwise the edges and their vertices under this map. Then the system  ${\cal M}$
is called a topological polygon.

Each polygon of ${\cal M}$ can be given a coherent orientation
of its edges \cite{SE} p. 132, \cite{MU} pp. 26-33. 
First one chooses an individual orientation on each 
edge, consistent 
with the topological identification. Then one chooses a single orientation 
on any consecutive sequence of edges forming a polygon. 
The orientation of the polygon
induces an orientation on  the edges belonging to it. 
In relation to the individual orientation it is marked by an exponent
$\pm 1$.

If the orientations of all polygons of ${\cal M}$ can be chosen such that
any edge gets two opposite induced orientations, 
${\cal M}$ is called orientable, otherwise nonorientable.

By $|V|, |E|, |F|$ denote the number of vertices, edges and faces
of a polygon . Then the Euler characteristics $\chi$ is defined by
\begin{equation}
\label{g1}
\chi= |V|-|E|+|F|.
\end{equation}
A topological polygon can be simplified into a fundamental polygon
without change of the orientability and characteristics. These
properties  uniquely classify  the compact surfaces.
The genus $g$ of the surface ${\cal S}$ is determined by the 
Euler characteristics and the orientability:
\begin{eqnarray}
\label{g2}
{\cal S}\; {\rm orientable}&:&\; \chi({\cal S})= 2- 2g({\cal S}),\nonumber \\
{\cal S}\; {\rm nonorientable}&:&\; \chi({\cal S})= 2- g({\cal S}).
\end{eqnarray}

In Table 1 we give from \cite{SE} the topological data of the sphere
$S^2$, the projective space $P^2$, the torus $T$, and Klein's bottle $K$.
The projective sphere can be described by closing the boundary of 
a M\"obius strip with a cross cap.
In Figs. 1,2 we sketch the fundamental polygons of these four surfaces.
The non-orientable surface $C^2$ with $\chi=-1,\; g=3$ 
will appear in section 5 as an embedding of the simplex. $C^2$ has
three cross-caps.
\vspace{0.2cm}

{\bf Table 1}: Euler characteristics $\chi$ and genus $g$ of fundamental polygons
for the sphere
$S^2$, the projective space $P^2$, the torus $T$, Klein's bottle $K$,
and a non-orientable surface $C$.
\vspace{0.2cm}

$$
\begin{array}{lllllll} \hline
{\rm Name}:&S^2&P^2& T &  K&C^2\\ \hline
|V|        &  2&  1&  1&  1&1\\
|E|        &  1&  1&  2&  2&3\\
|F|        &  1&  1&  1&  1&1\\ 
\chi       &  2&  1&  0&  0&-1\\
{\rm Genus\; g}&  0&  1&  1&  2&3\\
{\rm Orientable} &yes&no&yes&no&no\\ \hline
\end{array}
$$
\vspace{0.2cm}

In Figs. 1 and 2 we sketch the fundamental polygons for the
first four surfaces of Table 1.

\begin{center}
\input fupoSP 
\end{center}

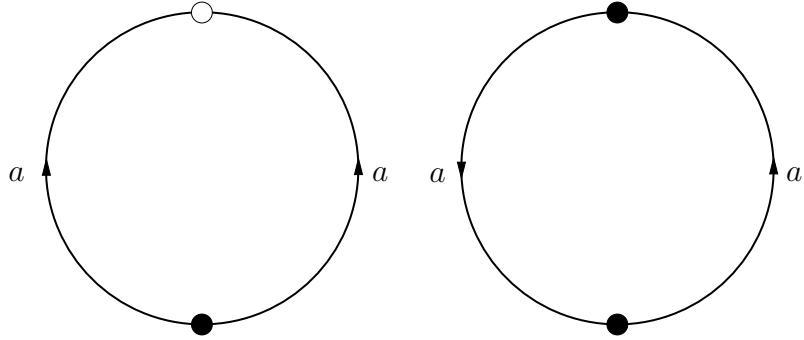
Fig. 1. Fundamental polygons for the sphere $S^2$ with two vertices, 
one edge $a$ and one face (left),  and for the projective
space $P^2$ with one vertex, one edge $a$ and one face (right).
The arrows determine the gluing of edges.
\vspace{0.2cm}

\begin{center}
\input fupoTK 
\end{center}

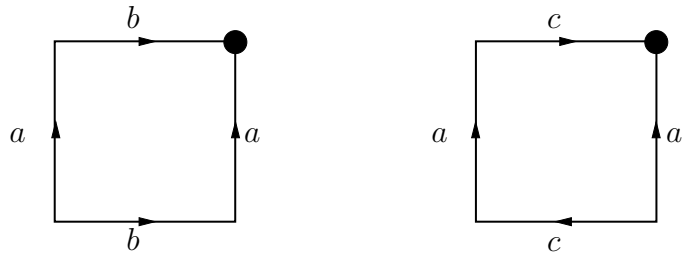
Fig. 2. Fundamental polygons for the torus $T$ with one vertex, 
two edges $a,b$ and one face (left), and 
Klein's bottle $K$ with one vertex, two edges $a,c$ and one face (right).
\vspace{1cm}

We shall use in particular the sphere $S^2$ and the projective space 
$P^2$ for the embeddings. 

In Figs. 4 and 5
we follow Seifert and Threlfall \cite{SE} p. 10 and present  
the non-orientable surface $P^2$ and its tesselations
as  a sphere: The  upper und lower hemisphere are
projected to a single equatorial  plane containing the equator, and 
opposite points on the equator are identified.

\section{Embedding of a graph into a surface.}

For notions of graph theory we refer to
Biggs \cite{BI}.  A graph $\Gamma$ consists of two sets $V\Gamma,
E\Gamma$, called the vertices and edges of $\Gamma$ with elements 
$v,e$ respectively, 
and an incidence relation, that is, a subset of $V\Gamma \times E\Gamma$.
We require that  every edge is incident with two vertices, 
and no two edges are incident with the same pair of vertices, that
is, the graph $\Gamma$ has no loops. We follow the exposition 
of Schwark \cite{SCH} of results due to Bonnington and Little \cite{BO}.

We define an embedding of a graph $\Gamma$ into a surface. 
Let $\Gamma$ be a graph and ${\cal S}$ be a compact
2-manifold or surface ${\cal S}$. A drawing of $\Gamma$ on ${\cal S}$ is 
a map of the vertices and edges into points and curves on ${\cal S}$ 
which preserves the relation between edges and vertices and moreover
has no intersection of edge curves except in their common vertices. Define
the components of ${\cal S}$ after removal of  all edge curves as
the regions of the drawing. If all the regions of the drawing are
homeomorphic to the Euclidean plane $E^2$ we call the drawing an
embedding of $\Gamma$ into ${\cal S}$.

Given such an embedding of a graph, we may attach the Euler characteristics and the 
genus $\chi, g$ to the embedded graph. Given a graph, its embedding
into a compact surface is not unique. We speak of a minimal embedding if
the  genus $g$ of the embedding surface is minimal.

We now follow \cite{SCH} and describe the steps which lead to embeddings of a given 
graph into surfaces. A graph $\Gamma$ can be reconstructed by incidences from
its set of edges as follows. Start from  a set  
$E, |E|$ and define on $E$
$X=: E \times
\langle -1,1\rangle,\; |X| = 2|E|$ with elements $x \in X$.
The edge set $E=E(X)$ in terms of $X$ is defined as
the  set of (unordered) pairs  $e:= ((e,-1), (e,1)) \in X \times X$. 
The pair $((e,-1), (e,1))$  forms the negative and positive ends 
\cite{BI} p. 24 of
the edge $e$. Choose the  set $V(X)$ as a partition $V$ of $X$ 
into subsets $v_{\alpha},
v_{\beta}, \ldots $ such that the subset 
$v_{\beta}$ reproduces the incidences of the ends of edges at the vertex 
$v_{\beta} \in V(\Gamma) \subset \Gamma$. It follows that from $(X,E,V)$ 
we can reconstruct $\Gamma$.

Now choose two bijections $\pi, \phi$  of $X$ as follows:
The orbits on $X$ under $\pi$ are the sets of ends for any 
vertex. Therefore $\pi$ is a unique involutive bijection which interchanges
the negative and positive ends on all the edges.
The bijection $\phi$ is chosen such that the orbits on $X$  under $\phi$ 
are all the vertex subsets $v_{\beta} \in X$. 
The choice of $\phi$ is not unique and becomes part of the classification 
of  embeddings of $\Gamma$.  Given $(X,E,V)$,
any  possible choice of $\phi$ provides
an order among the edges incident with any fixed vertex.  
Define a map ${\cal M}$ as a triple $((X,E,V),\pi,\phi)$
with a fixed choice of $\phi$.

From a map ${\cal M}= ((X,E,V),\pi,\phi)$ with underlying graph $\Gamma$,
we now assign a signature $i$ to its edges. We form $X \times 
\langle -1, 1\rangle$ with elements $(x,-1),(x,+1)$. The signature 
of an edge is a function $i(e)= \pm 1$. For $(x, \pm 1)\in e$ we also
assign $i((x,\pm 1))= i(e(x))=i(e)$. A map whose edges are equipped with a 
signature is denoted as ${\cal M}^i$.
For any element $(x,k) \in X \times 
\langle -1, 1\rangle$ and given signature we define \cite{SCH} an operator
\begin{equation}
\label{g101}
\Pi: (x,k)\rightarrow  \Pi(x,k) = (\phi^{ki(e(x))}\pi x, ki(e(x)))
\end{equation}  
$\Pi$ can be shown \cite{SCH} to be a bijection of $X \times 
\langle -1, 1\rangle$. Moreover its
orbits by construction pick one and only one point from each edge. 
These orbits therefore   define  a sequence of consecutive edges from
the graph $\Gamma$. Each orbit of edges under $\Pi$
determines a face of an embedding of the map ${\cal M}^i$ into
a surface. If there are $|F|$ orbits, the Euler characteristics of this
surface is obtained by application of  eq. \ref{g1}.

The orientability of an embedding can be decided as described in section 3.
The sets of consecutive edges become the orbits of edges 
from  the map eq. \ref{g101}.
Then eq. \ref{g2} can be used to find the genus $g$.

The topology of a surface ${\cal S}$  is further characterized by their
homotopy and homology groups. These topological data must also characterize
the embedded cell complexes.  We refer to \cite{SE} for details on these 
groups.

\section{Simplex embedding into compact surfaces.}

We demonstrate the algorithm of Schwark \cite{SCH} on  the 
graph $\Gamma$ for a simplex. The simplex corresponds to the $6j$ symbol. It is the first
non-trivial spin network and is used  in the geometric approach
to quantum gravity due to    Ponzano and Regge \cite{PON}.

A planar graph $\Gamma$ for the simplex is shown in the last entry
of Fig. 3. We label the edges by the letters $a,b,c,d,e,f$.
To each edge we attach first an initial and a final point $x$.
These points fall into $4$ vertex sets.  
To each point we assign  $k= \pm 1$. This gives the set 
$X \times \langle 1,-1\rangle$ of $24$ signed
points $(x,s)$. The bijection $\pi$ interchanges the initial and
final point on any vertex without changing the sign. 
The bijection $\phi$ must have the vertex sets of points as its
orbits. Therefore for the simplex, and more generally for any graph
of degree $3$ at any vertex, $\phi$ is the direct product $(Z/3Z)^4$
of  $4$ cyclic
groups of order $3$  at the four vertices. Our first choice $\phi=\phi_1$ 
for the simplex
is the product of counterclockwise cycles at all $4$ vertices.
Next we choose the signature $i= \pm 1$ on the edges.
We consider after one another signatures with $0, \ldots , 6$ negative
signs. For choosen signature we start the Schwark algorithm and
determine the orbits on $X \times \langle 1,-1\rangle$ under the
bijection $\Pi$. Our choices of signatures are given in {\bf Table 2}
for each running number $m$ in  alphabetic order $(a,b,c,d,e,f)$
of the edges.
Each orbit of points determines an orbit of 
edges, but any edge appears twice in different orientation.
In Fig. 10 we draw the orbits of edges as continuous lines running
parallel to a sequence of edges of the graph $\Gamma$. The number of edge 
orbits gives the number $|F|$ of faces of the embedding. Together
with $|V|=4, |E|=6$ this determines the Euler characteristics $\chi$
eq. \ref{g1} and from eq. \ref{g2} the genus $g$. All results for
our choice of $\phi$ are summarized in {\bf Table 2}. Observe that
any edge of $\Gamma$ appears exactly twice in the orbits. This property follows
from the assumed closure property of the embeddings 
which excludes edge lines containing
points on a boundary.
 
To check the orientability we choose the individual orientations on
the $6$ edges as shown by white arrows in the last entry of Fig. 3. The induced 
orientations on orbits of edges  are then marked
by exponents $\pm 1$ as given in {\bf Table 2}. In Fig. 3 a chosen 
orientation on polygon faces 
is marked by black arrows. All induced orientations on faces are 
marked by their exponents in {\bf Table 2}.

\begin{center}
\input simplex
\end{center}

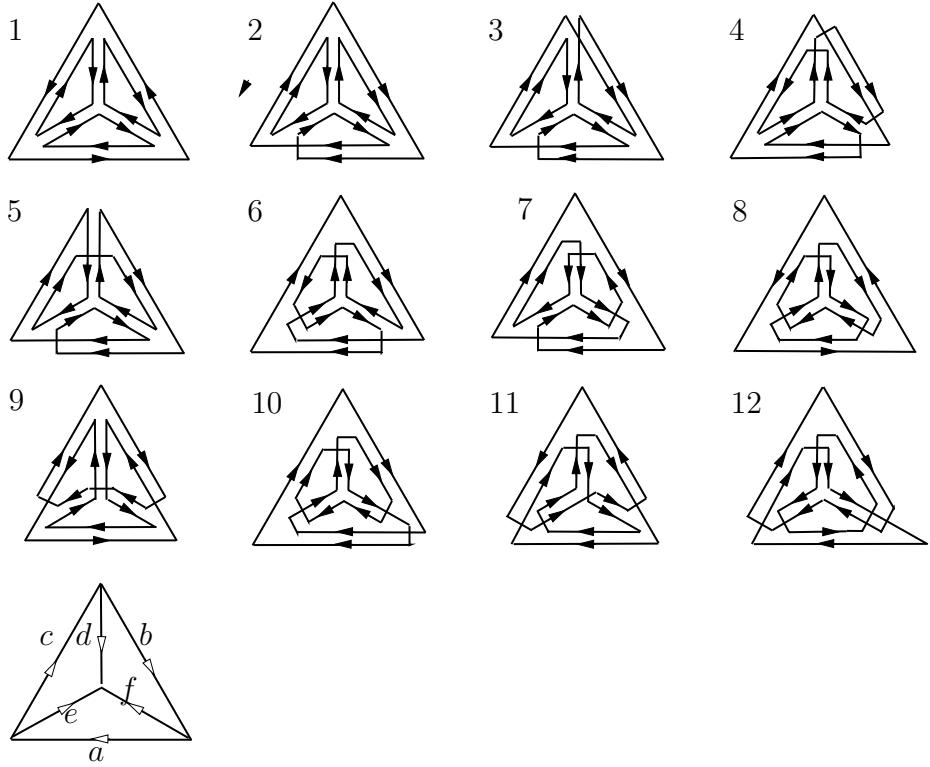
Fig. 3. $12$ embeddings of the simplex graph $\Gamma$ into compact surfaces. 
The last drawing gives 
the notation for edges of $\Gamma$ and their individual orientation 
by white arrows. Any face of the $12$  embeddings is drawn as 
a continuous line running piecewise parallel to an orbit of
edges of $\Gamma$ and oriented by the black arrows. 
Any edge appears twice on these lines (closure!). 
Crossings of lines do not imply intersections. The topological data, the 
orbits of edges which form faces, and the induced orientations  are listed in 
{\bf Table 2}. 
\vspace{0.2cm}

{\bf Table 2}: 
12 embeddings  of the simplex into compact surfaces.
The columns give for $m=1,\ldots, 12$ the signature $i$, the number of faces $|F|$, the 
characteristics $\chi$, the genus $g$, 
a symbol for ${\cal S}$ and its orientability $y/no$, and
orbits of  edges for faces with exponents for their induced orientation.

\begin{tabular}{llllllllllll}
$m$&i&&&&&& $|F|$& $\chi$&g&
sur/or& orbits of edges, ind. or.\\
1 & +&+&+&+&+&+& 4& 2&0&$S^2$/y
&$a^{-1}b^{-1}c^{-1}|aef^{-1}|bfd^{-1}|cde^{-1}$\\
2 & -&+&+&+&+&+& 3& 1&1&$P^2$/no
&$cde^{-1}|bfd^{-1}|acbaef^{-1}$\\
3 & -&-&+&+&+&+& 2& 0&2&$K$/no
&$c^{-1}ed^{-1}|acbfd^{-1}baef^{-1}$\\
4 & -&+&+&-&+&+& 2& 0&2&$K$/no
&$acbaef^{-1}|bfd^{-1}c^{-1}ed^{-1}$\\
5 & -&-&-&+&+&+& 1&-1&3&$C^2$/no
&$acde^{-1}cbfd^{-1}baef^{-1}$\\
6 & -&+&+&-&-&+& 1&-1&3&$C^2$/no
&$acbaed^{-1}bfd^{-1}c^{-1}ef^{-1}$\\
7 & -&+&+&-&+&-& 1&-1&3&$C^2$/no
&$acbaef^{-1}b^{-1}de^{-1}cdf^{-1}$\\
8 & +&+&+&-&-&-& 2& 0&1&$T$/y
&$a^{-1}b^{-1}c^{-1}|aed^{-1}bfe^{-1}cdf^{-1}$\\
9 & +&-&-&+&-&-& 2& 0&1&$T$/y
&$aed^{-1}c^{-1}a^{-1}b^{-1}df^{-1}|fe^{-1}cb$\\
10 & -&+&+&-&-&-&1&-1&3&$C^2$/no
&$acbaed^{-1}bfe^{-1}cdf^{-1}$\\
11& +&-&-&-&-&-& 2& 0&2&$K$/no
&$aed^{-1}bacdf^{-1}|b^{-1}c^{-1}ef^{-1}$\\
12& -&-&-&-&-&-& 3& 1&1&$P^2$/no
&$acdf^{-1}|a^{-1}b^{-1}de^{-1}|b^{-1}c^{-1}ef^{-1}$\\
\end{tabular}
\vspace{0.2cm}

The orientability with the criteria given in section 3 can be easily 
decided by inspection of the drawings in Fig. 3: Only in the drawings
with numbers $(1,8,9)$ is it possible to give each edge two opposite
induced orientations. In all other cases, allowing for the inversion 
of the orientation on faces, no induced orientation  can match all the pairs of
edges with opposite arrows.
The topological polygons for $S^2, P^2, T, K$ are given in section 3.
They would result from the embedded graphs by applying the topological
reduction algorithm given in \cite{SE} pp. 135-140.
{\bf Table 2} does not exhaust the simplex embeddings since we made 
only one choice of $\phi$. A straightforward but lengthy computation 
would yield all possible embeddings. 

There is a single embedding $(1)$ of minimal genus $g=0$ into the sphere
$S^2$, and there are two more orientable embeddings $(8,9)$ into the torus $T$. 
Two non-orientable embeddings $(2,12)$ go into $P^2$, 
three $(3,4,11)$ into K.
The four non-orientable embeddings $(5,6,7,10)$ have ${\cal S}=C^2$ 
of genus $g=3$. $C^2$ has $3$ cross-caps, and  there is a mirror 
pair $(6,7)$.

We now consider the dualization $\Gamma \rightarrow \Gamma^*$, 
compare section 7,  of the simplex $\Gamma \subset {\cal S}$
in the sense of algebraic topology \cite{MU}. It depends
on the chosen embedding. Under dualization of a 2-skeleton, 
vertices are interchanged 
with faces, and edges go into edges.
The topological data $\chi, g$ of the embedding in each case stay the same. 

The numbers of duals  are 
\begin{equation}
\label{g20}
|V^*|=|F|,\; |E^*|=|E|,\;|F^*|=|V|.
\end{equation}
and for the simplex from {\bf Table 2} range as 
$|V^*|=4,3,2,1,\; |E^*|=6,\; |F^*|=4$.
All the  cell complexes dual to the simplex are distinct triangulations 
of their compact surfaces and obey the rule $2|E^*|=3|F^*|$
typical for these. 
For the embedding $(1)$ of  minimal genus $g=0$ into $S^2$, the dual 
$\Gamma^*$ is the second simplex
often considered in relation with the $6j$ symbol.

\section{ Graphs $\Gamma(3nj)$ embedded into compact surfaces.}

The graphs $\Gamma=\Gamma (3nj)$ are characterized 
by the properties \cite{YU}:
\\ ($\Gamma$1): Each graph $\Gamma$ is connected and has $|E|=3n$ edges,
\\ ($\Gamma$2): The local degree at any vertex is three,
\\ ($\Gamma$3): The graph $\Gamma$ cannot be separated into two disjoint parts 
by cutting less than four lines.

Yutsis et al. \cite{YU}  describe in particular 
$3nj$ symbols and graphs  of a first and a second type. Moreover 
they give on \cite{YU} pp. 65-70 
methods by which new $3nj$ invariants $\Gamma'$ can be recursively constructed:
They  are generated from a known $\Gamma$  by 
putting two additional vertices on any two lines and connecting them.
From the Schwark algorithm we must expect for these graphs $\Gamma'$
embeddings into surfaces ${\cal S}$ of increasing genus $g$.

In what follows we shall consider only the first and second type of 
$3nj$ symbols and for short denote them by $3nj(1),\, 3nj(2)$. 
In \cite{YU} one finds planar graphs for these symbols.
The graphs for symbols of the second type can be redrawn in the plane
without crossing of edges.  
The planar graphs for symbols of the first type differ 
from the ones of the second type by a twist of two edges. 
Therefore they cannot be drawn in the plane without a crossing of edges.
We recall: A graph $\Gamma$ which cannot be drawn on the plane
without crossing of edges cannot be embedded into $S^2$ without 
crossing of edges. 
If a graph can be embedded into $S^2$ without crossing, it is topologically
equivalent to a convex polytope. But any convex polytope can be represented
by a Schlegel diagram on the plane without crossing.
It follows that the $3nj$ graphs of the first type
cannot be embedded into the sphere and has minimal embedding into $P^2$. 
The analysis of the simplex given in section 6 shows that
there will in general be a wide variety of embeddings. From 
Yutsis \cite{YU} we know that the number of $nj$-symbols
with different graphs will increase with $n$. The examples
of the simplex $6j$ and of the $9j$ symbol show that we must
expect embeddings into compact surfaces of increasing genus $g$.

Our results on $3nj$ graphs is obtained by a direct construction: 

{\bf 1 Prop}: The $\Gamma(3nj)$ graphs of the first type admit
polygonal embedding into the non-orientable projective space $P^2$ of 
minimal genus $g=1$. The minimal property results from the remarks made 
above. The topological data are listed in Table 3.
Examples are given in Figs. 4 and 5 for the
$9j(1)$ and $12j(1)$ symbols. Recall that objects in opposite position
must be identified.\\ 
The embedding of the $9j(1)$ graph into $P^2$ was already given by 
Ponzano and Regge \cite{PON}.

{\bf 2 Prop}: The $\Gamma(3nj)$ graphs of the second type admit a polygonal embedding
into the orientable sphere $S^2$ of minimal genus $0$. 
The topological data are listed in Table 4.
The example of $12j(2)$ is shown in Fig. 6.
\vspace{0.2cm}

{\bf Table 3}: Polygonal embedding for $\Gamma(3nj)$ graphs of first type.

\vspace{0.2cm}

$$
\begin{array}{lllll} \hline
{\rm Name}    & 9j & 12j& 15j& 3nj\\ \hline
|V|=|F^*|       &  6 &   8&  10&  2n\\
|E|=|E^*|     &  9 &  12&  15&  3n\\
|F|=|V^*|       &  4 &   5&   6& n+1\\ 
\chi    &  1 &   1&   1&   1\\
{\rm Genus\; g}   &  1 &   1&   1&   1\\
{\rm Surface} & P^2& P^2& P^2&P^2\\ \hline
\end{array}
$$
\vspace{0.2cm}

{\bf Table 4}: Polygonal embedding for $\Gamma(3nj)$ graphs of second type.
The $\Gamma(9j)$ graph is separable on $3$ edges, \cite{YU}.
\vspace{0.2cm}

$$
\begin{array}{llllll} \hline
{\rm Name}      &6j  & 9j  & 12j& 15j& 3nj\\ \hline
|V|=|F^*|       & 4  & 6   &   8&  10&  2n\\
|E|=|E^*|       & 6  & 9   &  12&  15&  3n\\
|F|=|V^*|       & 4  & 5   &   6&   7& n+2\\ 
\chi            & 2  & 2   &   2&   2&   2\\
{\rm Genus\; g} & 0  & 0   &   0&   0&   0\\
{\rm Surface}   & S^2& S^2 & S^2& S^2& S^2\\ \hline
\end{array}
$$
\vspace{0.2cm}

\begin{center}
\input 9j
\end{center}

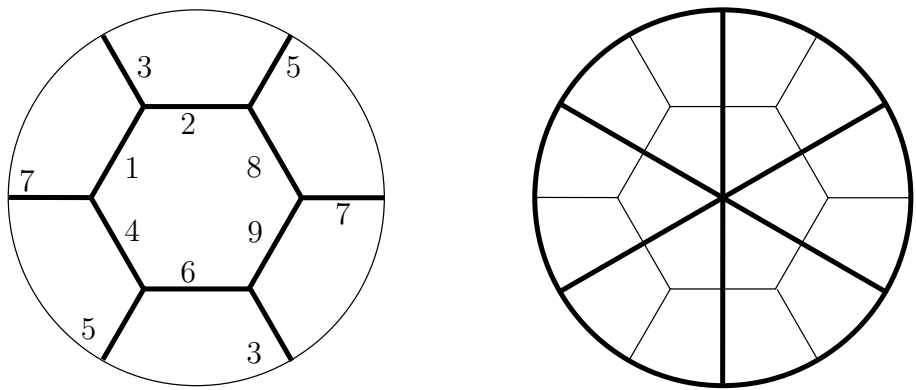
Fig. 4. The graphs of the $9j$ symbol of the first type embedded
into $P^2$.
A planar drawing is given in \cite{YU} Fig. 18.1.
Left: $2$-skeleton (heavy lines)  with $1$ hexagon 
and $3$ quadrangles. Right: Dual $2$-skeleton (heavy lines)  with $6$ triangles.

\vspace{0.2cm}

\begin{center}
\input 12j
\end{center}

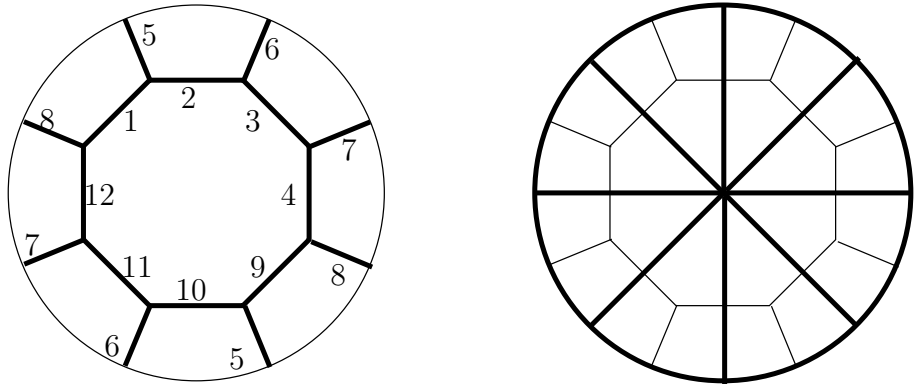
Fig. 5. The graphs of the $12j$ symbol of the first type embedded
into $P^2$. Left: $2$-skeleton (heavy lines) with $1$ octagon 
and $4$ quadrangles. 
A planar drawing is given in \cite{YU} Fig. 19.1a.
Right: Dual $2$-skeleton (heavy lines) with $8$ triangles.

\vspace{0.2cm}

\begin{center}
\input 12j2
\end{center}

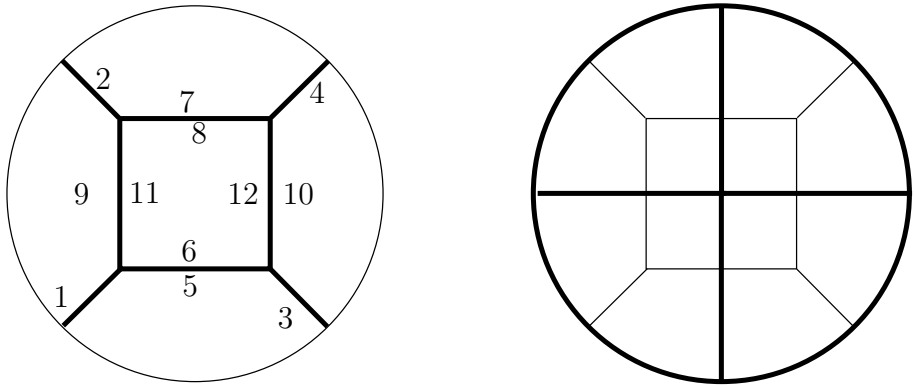
Fig. 6. The graphs of the $12j$ symbol of the second type embedded
into $S^2$. Left: cubic $2$-skeleton (heavy lines) with $6$ quadrangles. 
The edges $6,8,11,12$ are on top, the edges $5,7,9,10$ below  the 
equator. A planar drawing is given in \cite{YU} Fig. 19.2.
Right: Dual octahedral $2$-skeleton (heavy lines) with $8$ triangles.
\vspace{0.2cm}

For short we use integers $i$ to denote the entries $j_i$ of the $3nj$ symbols.
The $9j$ symbol of the first type with graph shown in Fig. 3 is in the notation of 
\cite{YU} eq. (18.2) 
\begin{equation}
\label{g51}
9j(1):=
\left[
\begin{array}{lll} 
1&2&3\\
4&5&6\\
7&8&9
\end{array}
\right]
\end{equation}
The $12j$ symbols of the first and second type with graphs shown 
in Figs. 4 and 5   in the notation 
of \cite{YU} eq. (19.1) and eq.(19.5) respectively are
\begin{equation}
\label{g52}
12j(1):=
\left[
\begin{array}{llllllll} 
1& &2& &3& &4&\\
 &5& &6& &7& &8\\
9& &10& &11& &12&
\end{array}
\right],\;
12j(2):=
\left[
\begin{array}{llll} 
1&2&3&4\\
5&6&7&8\\
9&10&11&12
\end{array}
\right].
\end{equation}

The embeddings of the two families of $3nj$ graphs into
$P^2$ and $S^2$ are of minimal genus. 
An important property of the $3nj$ symbols is their symmetry \cite{YU} 
under certain permutations of the entries $j$.
Symmetries keep the numerical value of the corresponding $3nj$ invariant
up to a sign.  All of them
provide automorphisms of their graphs. A discussion of these symmetries
in terms of the surface embeddings is possible but will not be given 
here.

From the compact surface embeddings of the $3nj$ graphs one could pass 
to other embeddings. Any compact surface ${\cal S}$ has a simply connected 
universal covering ${\cal U}({\cal S})$. An embedding 
$\Gamma \subset {\cal S}$ may be extended to  ${\cal U}({\cal S})$. 
An embedding into $E^3$ is obvious for $ \Gamma  \subset {\cal S}=S^2$. 
For the higher embeddings 
of the projective space $P^2$ into $E^3$ see
\cite{HI} pp. 313-329 and into $E^4$ pp. 340-342.

\section{Dualization of $3nj$ graphs.}

The concept of dualization,  \cite{MU} pp. 367-447, applies to general 
cell complexes $C$ of dimension $d$ and 
their skeletons or subcomplexes of dimension $m,\; 0 \leq m \leq d$. 
For general dimension $d$ of the complex, its dual objects in $C^*$ have
complementary dimension $(m, d-m)\; m=0,\ldots, d$.
Duality therefore depends on the dimension of the cell complex.
For the present  cell complexes embedded into surfaces  
we have $d=2$ and so faces are dual to points and edges dual to edges.

A graph when imbedded into a surface ${\cal S}$ extends the graph
by its faces to a  
2-skeleton whose 1-skeleton is the graph. For simplicity we shall denote
this embedded cell complex again by $\Gamma$.
The dual 2-skeleton $\Gamma^*$ on the same surface ${\cal S}$
may be obtained as
follows: As vertices of $\Gamma^*$ choose interior points from
each face of $\Gamma$. Connect by a dual edge each pair of interior points
from pairs of polygons which share an edge. Then the dual edges must close
around any vertex of $\Gamma$. Assign a dual face to each such set of 
closed dual vertices. Clearly  the dual graph is embedded  into the
same surface ${\cal S}$. Euler characteristics $\chi$, genus $g$ and 
orientability are unchanged. The interchange of boundaries follows eq. \ref{g20}.
All dual complexes are triangulations and obey $2|E^*|=3|F^*|$.
The new edges of the dual 2-skeletons could
carry the same values of $j$ as the old ones. 

For the $3nj$ graphs considered we find for dualization: 

{\bf 3 Prop}: The 
dual 2-skeletons $\Gamma^*(3nj)$ for the first and second type of 
$\Gamma(3nj)$ graphs 
with minimal embedding admit
triangulations, non-orientable on
the projective space $P^2$, and orientable on 
the sphere $S^2$  respectively. 

The embeddings and triangulations for the first type of $\Gamma(3nj)$ graphs 
into $P^2$ should be handled
with care: Once the upper and lower hemisphere of the orientable $S^2$ 
are identified
on a  single non-orientable hemisphere, there is no volume left in between.
For higher embeddings of $P^2$ compare \cite{HI} as quoted above.

The 
number of vertices, edges and faces and the topological data for the
duals 
are obtained  by interchange of the entries
for vertices and for faces, see Tables 3 and 4.
In Figs. 4, 5 and 6 we show examples of  polygonal embeddings on the left and their
dual triangulations on the right. 
The generalizations to $n>4$ are straightforward.

Non-minimal embeddings with higher genus $g$ could be constructed by the Schwark
algorithm with the methods demonstrated in section 5 for the simplex.

Finally we mention an alternative elegant graphical representation of invariants and
recoupling coefficients due to Fano and Racah \cite{FA} pp. 156-158,
see also \cite{PON} p. 32. 
Each triple of
coupled vectors  in the Fano-Racah-graphs  labels $3$ points
on a straight line. We do not see an easy passage  from the Fano-Racah
graphs to the topology and dualization in question.

\section{Conclusion.}

The construction of maps due to 
\cite{BO} and \cite{SCH} allows to obtain all embeddings of spin networks
$\Gamma$ into compact surfaces ${\cal S}$. The embeddings give insight into
a variety of topological data as $\chi, g$, orientability, homotopy and
homology, duality and  triangulation.
For two types of $\Gamma(3nj)$ networks with general $n$, there are 
embeddings of  minimal genus into  
the sphere and the projective space. 
The dual 2-skeletons $\Gamma^*(3nj)$ are triangulations.

We take up the topics of the introduction and draw some conclusion 
on future applications of spin networks, in particular  in the spirit of 
the initial approach of Regge \cite{REG2} and Ponzano and Regge \cite{REG}.
The results reached in this field have been summarized by
Regge in \cite{REG}. In view of the variety of distinct embeddings,
topologies  and dualizations 
which we have demonstrated in sections 5 and 6 one could ask
for some guidelines on the  topology of spin networks.\\ 

(1) In our opinion, the embedding of spin networks 
into surfaces ${\cal S}$ and the 
topological data must 
be implemented  before dualization, triangulation,
and before any more advanced geometry and physics of spin networks.\\
(2) The simplest choice of embeddings 
would be to admit them  only into ${\cal S}= S^2$. These correspond
to convex polyhedra in Euclidean space $E^3$, and would allow for
standard geometric tools. All $3nj$ graphs of the second type 
admit this $g=0$ embedding.\\
(3) The next possible choice would be
embeddings of minimal genus. All 
$3nj$ graphs of the first type, beginning with the $9j$ graph
discussed already in \cite{PON},
admit the embedding into $P^2$ with minimal genus $g=1$.
Compact embeddings of other characteristics and genus are 
evident from the embedding of the simplex, section 5.
The projective space $P^2$ can be seen as a sphere with the identification
of opposite points. But the  dual triangulation of $P^2$ is not orientable,
has no
straight-forward interpretation as
the boundary of a polytope, and so the geometric view of faces and volumes
as suggested in \cite{PON} requires further topological qualification.\\
(4) A more general point of view is suggested
by modern cosmology. The study of varieties of
cosmologies with non-standard topology is in progress \cite{LA}, and
their tests from  astronomical evidence are under intense
study. Why should  discrete models of space not
display a similar  richness of topology as continuous ones?
A restriction to minimal embeddings of graphs may be only a 
first choice
on heuristic grounds. General embeddings of graphs $\Gamma$ 
into compact surfaces
${\cal S}$ with topological  data including
homotopy and homology, dualization and possible triangulation can
illuminate the way towards discrete geometries with richer  topology.\\
(5) Spin networks have been recently applied to quantum gravity 
\cite{CA}. The difference with other approaches discussed in 
\cite{CA} is that they offer new insight into the interpretation 
of space, a fundamental aspect of which is the correspondence 
between the combinatorial rules of the networks and the 
geometrical properties
of space. Also a very important consequence of spin networks is that all physical 
magnitudes are discrete, a property that can be used in the
quantization of general relativity \cite{LOL}.

\section*{Acknowledgments.}

The authors want to express their gratitude to Professor Bruno 
Gruber for the invitation to present these ideas in the Symposium
SYMMETRIES IN SCIENCES XII. One author (M. L.) 
gives thanks to the Ministerio de Ciencia y Tecnologia for financial support
under grant BFM 2000-0357.

\end{document}